# Simulation and Parameterization of Longitudinal Development in Extensive Air Showers for Different Hadronic Interaction Models

Kadhom F. Fadhel[1,2, a], A. A. Al-Rubaiee[2, b]

[1] Directorate General of Education in Diyala, Ministry of Education, Baghdad, Iraq

[2] Department of Physics, College of Science, Mustansiriyah University, Baghdad, Iraq

[a]kadhumfakhry@uomustansiriyah.edu.iq, [b]dr.rubaiee@uomustansiriyah.edu.iq

**Abstract**

The simulation analysis of the Extensive Air Showers (EAS) was executed by exploring the longitudinal development employing the AIRES system (version 19.04.00) for several hadronic interaction models (SIBYLL, QGSJET, and EPOS) for high energies. The simulation was performed for different high energies ($10^{17}$, $10^{18}$, and $10^{19}$) eV and two dissimilar primary particles, proton as well iron nuclei, with several zenith angles values ($0^o$, $10^o$, and $30^o$). The shower size of longitudinal development was parameterized using the sigmoidal function (Boltzmann model) and gave a new four parameters as functions of the primary energy between the energy extent ($10^{17}$-$10^{19}$) eV. The comparison among the acquired results data (the parameterized number of shower particles) along with the experimental results (Pierre Auger experiment) had offered a fascinating matching for the primary proton at the fixed primary energy $10^{19}$ eV for vertical EAS showers.

## 1. Introduction

The most considerable number of energetic particles scattered throughout the universe was classified as Ultra-High Energy Cosmic Rays (UHECRs). The study related to the cascades generating from their interactions with atmospheric nuclei will exhibit several magnitude orders greater than those obtained inside human-made colliders with a rare insight for interaction attributes of hadronic located in center-of-mass energies [1]. The collision between both an atmospheric nucleus and UHECRs launches in EAS of a certain secondary particles incrementing in the traversed air mass, typically described to an incline depth, $X$ [2]. Furthermore, EAS are a cascade of particles caused via the interact of a single superior energy primary cosmic ray, for example as (Fe, He, C, and p) with atmospheric nuclei comprising (O, N, and Ar). EAS enlarge in a complicated style as a combining between the electromagnetic cascades, hadronic and muonic cascades multi-particles creation [3, 4]. Extensive numerical simulations of air showers are required to be performed for extrapolating the characteristics of the primary cosmic rays which start them [5]. In 2005, J. Matthews used a simple semi-empirical model to develop the hadronic portion of air showers in a manner analogous to the Heitler splitting. Various characteristics of EAS are plainly exhibited with numerical predictions in good accord data with detailed Monte Carlo simulations. Results for energy reconstruction, muon and electron sizes and for the effects of the atomic number of the primary were clearly estimated [6]. In 2008, M. Unger presented a new method for the reconstruction of the longitudinal profile of EAS induced by ultra-high energy cosmic rays. This method employs directly the ionization energy deposit of the shower particles in the atmosphere. Furthermore, the extrapolation of the observed part of the profile with a Gaisser–

Hillas function is used and the total statistical uncertainty of shower parameters like total energy and shower maximum had been calculated [7]. Additionally, the longitudinal progress related to EAS is counting on the character of primary particle and energy particle [8]. Moreover, the charged particles number *N* into EAS as a function related to atmospheric depth is thoroughly linked to the energy and nature of the primary particle that perfectly can be examined via the AIRES code [9]. Thus, the simulation analysis uptake AIRES system was done for two types of primary particles (primary proton as well as iron nuclei) via the range of energy ($10^{17}$-$10^{19}$) eV for various zenith angles ($0^0$, $10^0$, and $30^0$). The longitudinal development parameterization that started in EAS was executed depending on the sigmoidal function referred by (Boltzmann model), and provide four parametric functions as functions of the primary energy.

## 2. Longitudinal Development in EAS

The Gaisser-Hillas Function will accurately parameterize the longitudinal EAS profile, refer to the growth of the charged particles number (indicates that shower size) along with the atmospheric depth *X*, (which is the actual depth cut inside atmosphere), [10]:

$$N(X) = N_{max}\left(\frac{X-X_o}{X_{max}-X_o}\right)^{(X_{max}-X_o)/\lambda} \exp(X_{max}-X)/\lambda, \tag{1}$$

where $N_{max}$ is maximum shower size; *X* the observation depth; $X_0$ is the depth related to the first interaction and $X_{max}$ is depth of the shower maximum [11]. This is a Gamma function type that automatically originates in cascade theory and can be employed to characterize unpredictable bounded variables. The overall main formula which describes the probability density function related to the gamma distribution is presented as in formula [12]:

$$f(X) = \left(\frac{X-u}{\lambda\Gamma(N)}\right)^{n-1} \exp{-(X-u)/\lambda}, \qquad X \geq u, \tag{2}$$

where *n*, is the shape parameter, *u*, is the location parameter and λ, is a distinctive length parameter. If $u = X_0$, we obtain [12]:

$$N(X) = N_0\left(\frac{X-X_o}{\lambda(N)}\right)^{n-1} \exp{-(X-X_0)/\lambda}, \quad n>1, \tag{3}$$

and

$$N_{max} = N_0(n-1)^{n-1}\exp{-(n-1)} \tag{4}$$
$$X_{max} = (n-1)\lambda + X_0 \tag{5}$$

The Gaisser – Hillas Function can be produced from the previous equations, however, is partially unlike pristine from which stated by Gaisser and Hillas [13]:

$$N(X) = N_{max}\exp(X_{max}/\lambda) - 1\left(\frac{X-X_o}{X_{max}-\lambda}\right)^{(X_{max}/\lambda)-1} \exp{-(X-X_0)/\lambda}, \tag{6}$$

## 3. Results and Discussion
### 3.1 AIRES Simulations

The specifics of shower evolution are extremely complicated to be completely delineated by uncomplicated analytical modelling. Monte Carlo simulation of interaction and transport of every single particle is needed to execute for accurate shower modelling evolution. Lately, Monte Carlo packages are employed for simulating EAS using AIRES (AIR shower Extended Simulation) system [10]. Various hadronic interaction models are utilized for these event generators, for example SIBYLL [14], QGSJET [15] and EPOS [16]. The air shower simulation programs consist of various interacting procedures that operate on a data set with a variable number of records, modifying its contests, increasing or decreasing its size accordingly with predetermined rules. The simulation engine of AIRES also possesses internal monitoring procedures that constantly check and record particles reaching the ground and/or passing across predetermined observing surfaces located between the ground and injection levels. Where the number of showers is determined and then the identity of the elementary particle is determined, as well as its energy that will interact with the atoms of the atmosphere. Then we define the name of the task, as well as the kinetic energy of electrons, muons, and gamma rays. Next, we define the thinning energy and the zenith angle, and then choose the observing levels for the array to be used. And finally, we define the name of the secondary particles resulting from the chain reaction. Figure 1 contains a schematic representation of the standard structure of the main AIRES simulation program [10]. The diffractive interactions possess a straight influence onto the shower progress. Also, that fact is clearly confirmed by graphing the number of charged particles versus the depth of atmosphere, at certain value of energies of $10^{17}$, $10^{18}$, and $10^{19}$ eV. The graphs were plotted depending on the data incoming from simulations executed via the AIRES system for assorted hadronic interaction models (SIBYLL, QGSJET and EPOS). The simulation was employed to investigate the production of primary particles (proton as well as iron nuclei) resulted from air showers within the range of primary energy ($10^{17}$-$10^{19}$) eV and explore the longitudinal growth of various hadronic interaction models created subsequent primary cosmic rays of the extremely high value of energy react with the atmosphere and organize overall correlated production data [4]. Whole simulation process is executed via the AIRES code which gained by utilizing the thinning factor $10^{-6}$ relative.

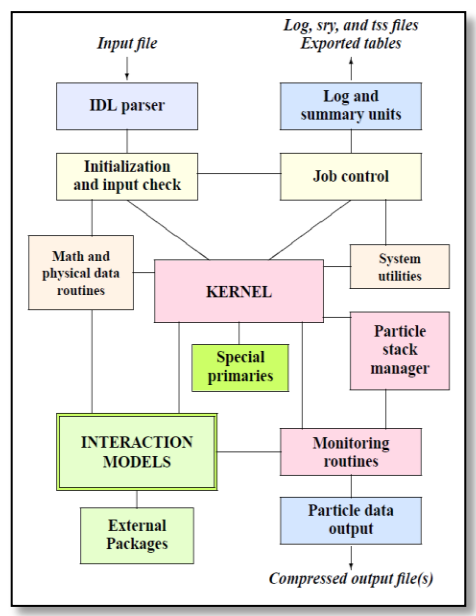

**Figure 1.** Structure of the AIRES main simulation

In figure 2 was displayed the dissimilarity between both the primary proton as well as the iron nuclei for simulating longitudinal progress at the primary energy $10^{19}$ eV and vertical EAS showers for different secondary particles, comprising electron and positron pair production and muons charged secondary particles.

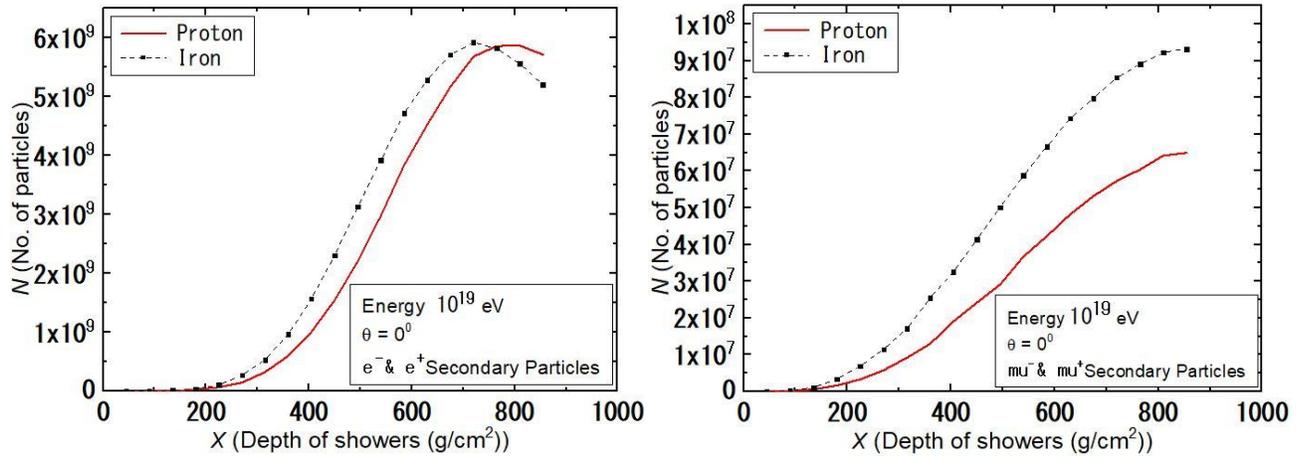

**Figure 2.** The longitudinal development which simulated via AIRES code at vertical showers for proton as well iron nuclei primary particles.

In figure 3 was shown the simulation of longitudinal development using AIRES system for different hadronic interaction models (SIBYLL, QGSJET, and EPOS) for pair of primary particles like proton likewise iron nuclei at a certain fixed energy $10^{18}$ eV for inclined zenith angle ($\theta = 10°$), for muons charged secondary particles.

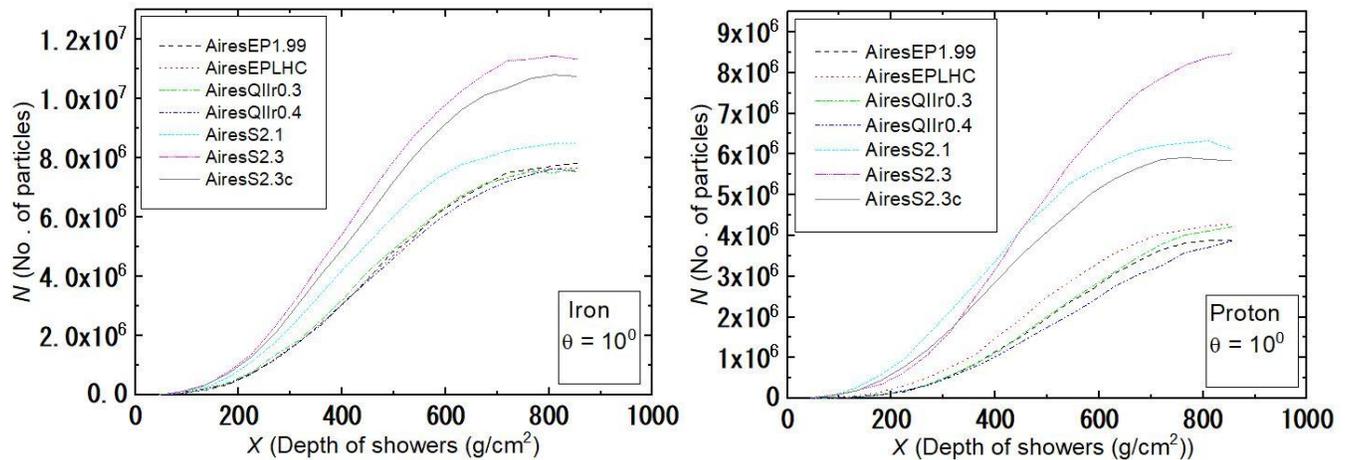

**Figure 3.** Comparison with different hadronic interaction models using AIRES system of longitudinal development at the primary energy $10^{18}$ eV for secondary muons.

## 3.2. Parameterization of longitudinal development

The Aires simulation was applied for primary proton as well as iron nuclei within the primary energy extent ($10^{17}$-$10^{19}$) eV and explores the longitudinal development for different hadronic models. The parameterization of longitudinal evolution for showers that started in EAS was executed relying on sigmoidal function (Boltzmann's model), which produced a novel four parameters for diverse primary particles. The coming formula can be represented this function as the following:

$$N(E) = \frac{\eta - \zeta_c}{1+ \exp\left(\frac{(x-\delta)}{\alpha}\right)} + \zeta_{c'} \tag{7}$$

where $N$ is the number of particles in (EAS) as a function of primary energy; $\eta$, $\zeta_c$, $\delta$, and $\alpha$ are the gained coefficients for the longitudinal development (tabulated in Table 1). These coefficients are obtained by fitting the AIRES results, which are given by the polynomial form:

$$K(E) = a_o + a_1(E/eV) + a_2(E/eV)^2 \tag{8}$$

where $K(E) = \eta, \zeta_{c'}, \delta, \alpha$ are parameters of Eq. (7) as a function of the primary energy and their coefficients $a_o, a_1$, and $a_2$ (listed in Table 1).

**Table 1.** Coefficients of the sigmoidal function (Boltzmann's model) (Eq. 7) that employed to parameterize the AIRES simulation for two primary particles within the energy range ($10^{17}$-$10^{19}$) eV and various zenith angles.

| Primary particle | Secondary particles | $K(E)$ eV | Coefficients | | |
|---|---|---|---|---|---|
| | | | $a_o$ | $a_1$ | $a_2$ |
| P | $e^-$ & $e^+$ | $\eta$ | $3.03 \times 10^5$ | $7.007 \times 10^{-12}$ | $-9.18 \times 10^{-31}$ |
| | | $\zeta_c$ | $-2.53 \times 10^6$ | $5.73 \times 10^{-10}$ | $3.25 \times 10^{-30}$ |
| | | $\delta$ | 360.83 | $6.47 \times 10^{-17}$ | $-4.7 \times 10^{-36}$ |
| | | $\alpha$ | 46.5 | $1.14 \times 10^{-17}$ | $-8.1 \times 10^{-37}$ |
| | mu$^-$ & mu$^+$ | $\eta$ | 5916.11 | $-4.93 \times 10^{-13}$ | $2.51 \times 10^{-32}$ |
| | | $\zeta_c$ | $6.23 \times 10^3$ | $9.92 \times 10^{-12}$ | $-2.92 \times 10^{-31}$ |
| | | $\delta$ | 350.56 | $7.24 \times 10^{-17}$ | $-5.52 \times 10^{-36}$ |
| | | $\alpha$ | 96.56 | $2.2 \times 10^{-17}$ | $-1.9 \times 10^{-36}$ |
| Fe | $e^-$ & $e^+$ | $\eta$ | $3.99 \times 10^5$ | $4.1 \times 10^{-12}$ | $-4.25 \times 10^{-31}$ |
| | | $\zeta_c$ | $-6.93 \times 10^6$ | $5.51 \times 10^{-10}$ | $2.42 \times 10^{-30}$ |
| | | $\delta$ | 338.23 | $8.78 \times 10^{-17}$ | $-7.38 \times 10^{-36}$ |
| | | $\alpha$ | 48.09 | $8.27 \times 10^{-18}$ | $-6.2 \times 10^{-37}$ |
| | mu$^-$ & mu$^+$ | $\eta$ | $-5.27 \times 10^4$ | $-4.06 \times 10^{-13}$ | $-5.09 \times 10^{-33}$ |
| | | $\zeta_c$ | $3.54 \times 10^5$ | $1.14 \times 10^{-11}$ | $-1.58 \times 10^{-31}$ |
| | | $\delta$ | 356.31 | $7.14 \times 10^{-17}$ | $-5.9 \times 10^{-36}$ |
| | | $\alpha$ | 112.24 | $-3.25 \times 10^{-18}$ | $4.73 \times 10^{-37}$ |

Figure 4 shows the parameterization as a function of primary energy of the shower size in EAS using sigmoidal function (Boltzmann's model) (Eq. 7) for a primary certain energy $10^{19}$ eV for two zenith angles ($0^0$ and $30^0$) for muons charged secondary particles.

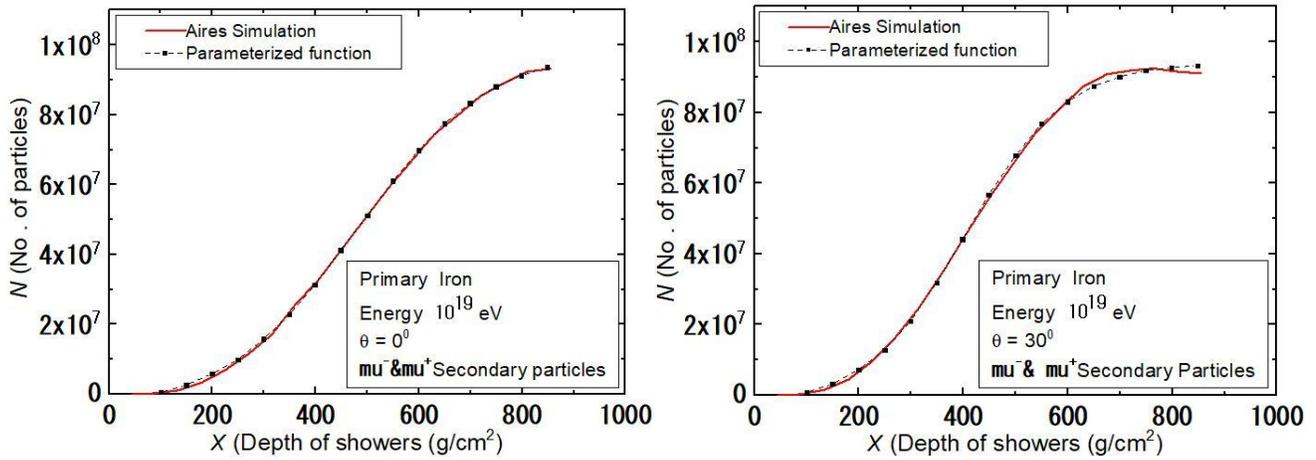

**Figure 4.** The difference between the simulation AIRES system and the results obtained using equation (7).

### 3.3. The comparison with Pierre Auger Obsevetory

Compared the parameterized longitudinal development shower size that was obtained using Eq. 7 (Sigmoidal Function) with experimental results for the Pierre Auger experiment [2, 17] which gave a good agreement for vertical EAS showers for the primary proton at fixed energy 1019 eV and various secondary particles. (a) electron and positron pair production (b) muons charged secondary particles.

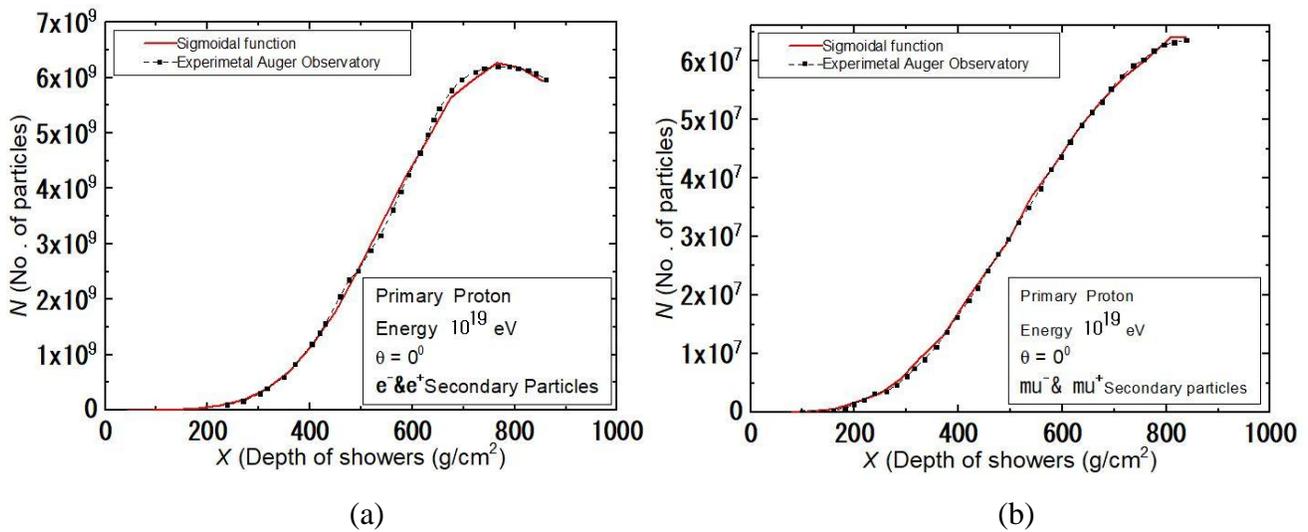

(a)  (b)

**Figure 5.** Comparison of the parameterized shower size of longitudinal development with the experimental data obtained by Pierre Auger Observatory for primary proton at the energy $10^{19}$ eV.

## 4. Conclusions

AIRES code was used for different high energy hadronic interaction models comprising (SIBYLL, QGSJET, and EPOS). Simulating the development of air showers in EAS performed for two primaries (proton and iron nuclei) for a certain energies values ($10^{17}$, $10^{18}$, and $10^{19}$) eV and various zenith angles like ($0^o$, $10^o$, and $30^o$). In the current work, an equation has been developed for the high-energy simulation process (Sigmoidal function) by making this equation depends on new four parameters. The longitudinal development parameterization of EAS showers was performed using the sigmoidal function (Boltzmann's model) that gave a four new parameters for different primary particles. Also, the shower size parameter was obtained using the results of AIRES simulation as a function of primary energy. The comparison of the parameterized longitudinal profile with that measured with the Pierre Auger experiment demonstrates the ability for identifying the primary particles and to determine their properties at the very high energies around the ankle region of cosmic ray spectrum. The main advantage of the given approach may give a possibility of longitudinal development analyzing for real events that detected in EAS arrays and mass composition and for reconstructing the primary cosmic ray energy spectrum.

**Acknowledgments**